\documentclass{millour}
\usepackage{graphicx}
\usepackage{hyperref}
\usepackage{natbib}

\TitreGlobal{What can the highest angular resolution bring to stellar astrophysics?}
\begin{document}

\title{Olivier~Chesneau's work on Novae}
\runningtitle{}
\newcommand{\folder}{.}


%
\author{F. Millour}\address{Laboratoire Lagrange, UMR7293,
  Universit\'e C\^ote d'Azur, Observatoire de la C\^ote d'Azur, CNRS,
  Bd. de l'Observatoire, 06300 Nice, France\\\email{fmillour@oca.eu}}
\author{E. Lagadec$^1$}
\begin{abstract}
Olivier~Chesneau founded a brand new field of observational
astrophysics with his attempts to resolve the novae expanding fireball
from the very first days of the explosion. With the images he could
get, he showed that novae do indeed explode in an aspherical way,
leading to a change of paradigm for the physics of these yet-poorly
understood catastrophic systems. He also set the stage for a new way
of estimating novae distances, by directly measuring the sky-size of
the fireball and comparing it with spectroscopic scales, taking into
account the tremendous effects of the fireball geometry.
\end{abstract}
\maketitle
\section{Introduction}

Novae are thought to be the progenitors of type-Ia supernova, the
standard candle for remote distance gauge in the Universe.

They are symbiotic stars, made of a giant star and a white dwarf
orbiting around each-other. These two close-by stars interact through
a Roche lobe overflow on the giant star side, feeding an accretion
disk on the white dwarf side. The accretion piles up hydrogen in large
quantity on the surface of the white dwarf, until the accreted mass is
sufficient to ignite again nuclear reactions. A thermonuclear
explosion occurs, blowing away most (but not all) of the material
accreted onto the white dwarf. The very faint stellar pair light is
quickly overcome by an extremely bright fireball, which soon engulfs
the whole system, and later relax in the interstellar medium in
timescales of months to years. Novae exhibit changes of up to 15
magnitudes within a few days. Once the relaxation is over, the process
of accretion / explosion cycles can start again, but the remaining
mass of hydrogen at each explosion piles up on the surface of the
white dwarf, accelerating the cycle every time up until a type-Ia
supernova occurs.

Novae remnants often show aspherical shapes, which were not
well-explained before, as the fireball was supposedly round in the
first moments of the explosion. Olivier Chesneau realized that he
could get the shape of the fireball in the very first days of the
explosion with the new high-angular resolution techniques he was using
in other fields \citep[see a summary of his research in the previous
  articles][]{Lagadec2015, Millour2015}. This would enable him to
trace the link between the spherical fireball and aspherical
remnant. Getting to such early moments mean having a sufficient
angular resolution of a few milli-arc-seconds, unreachable by
conventional telescopes, even equipped with adaptive optics. Only the
long-baselines of stellar interferometers can attain such a high
angular resolution.

O.~Chesneau found out that no multi-baselines studies of nova had been
done with interferometry, and that such observations could be a
revolution for the study of the environments of such objects. Time
series of interferometric observations enable one to gauge the angular
expansion of the fireball. Knowing its expansion velocity (with
simultaneous spectroscopy) can lead to a direct estimate of the
distance of the symbiotic system.

Of course, if the nova fireball is aspherical, or even more:
asymmetric, then a catastrophic bias in the distance estimate may
sneak in the whole process. Fortunately, such observations can also
enable the study of the geometry of the nova from the very early
stages of the explosion, and the monitoring of its evolution. His
dream was to be able to get both the geometry from the first blink of
the explosion, and to monitor the expansion of the fireball, making
accurate distance estimates of galactic novae slide from the
science-fiction to science side.

In addition, for dust-producing novae, the quantity of dust production
over time after the nova explosion can be accurately weighted.

\section[Novae]{Live formation of bipolar novae: RS Oph \& T Pyx}

O.~Chesneau started discussing with different novae specialists to
prepare an ESO proposal to observe novae with the VLTI. He estimated
that he should be able to observe a nova with AMBER every 2 years and
with MIDI every 4 years, as the targets needs to be closer than
$\sim$4 kpc to be large enough and bright enough for the VLTI. He
wrote a proposal with a few colleagues from Grasse, but it was
rejected. He nevertheless decided, together with Christian Pollas and
Alain Spang to monitor potential targets. The eruption of the
recurrent nova RS Oph in 2006 was a perfect opportunity for
O.~Chesneau to propose ESO DDT (Director Discretionary Time)
observations of a nova.

He was incredibly fast in writing down the proposal. The nova exploded
the $12^{\rm th}$ of February 2006, the DDT was basically ready the
$13^{\rm th}$ (late in the night...). The DDT was also accepted
extremely fast due to the accidental presence of the whole DDT board
at a meeting in ESO at that time. The observations were therefore
carried out just 5.5 days after the outburst, the $18^{\rm th}$, and a
first set of high-quality AMBER data was delivered. Unfortunately, the
following sets were of much lower quality, making them unusable in
practice.

\begin{figure}
\centering
\frame{\includegraphics[width=4cm]{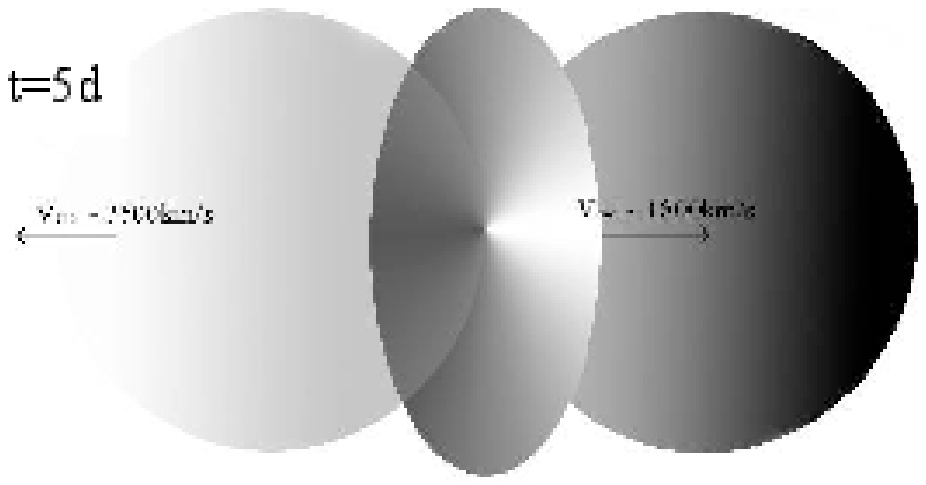}}
\frame{\includegraphics[width=4cm]{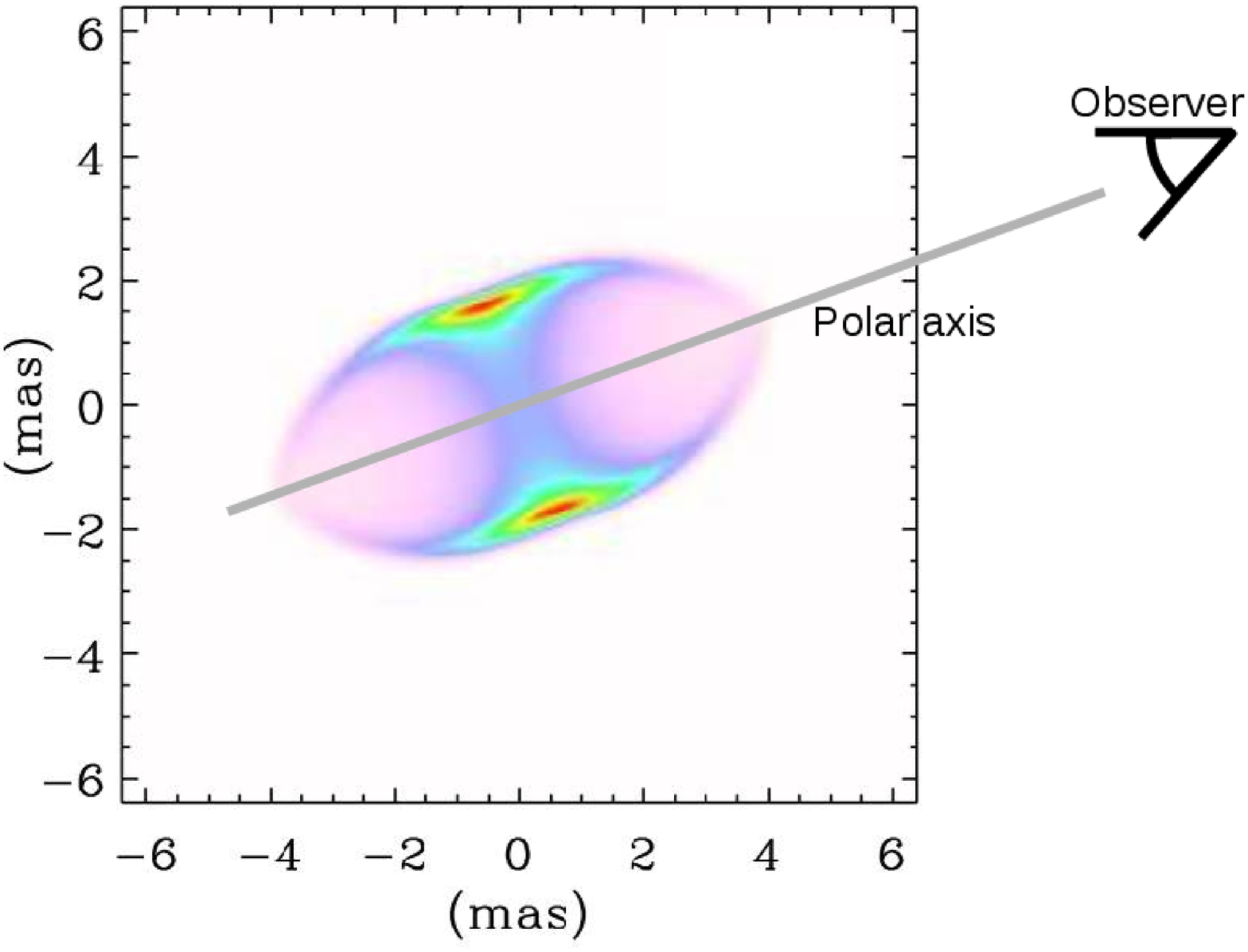}}
\frame{\includegraphics[width=4cm]{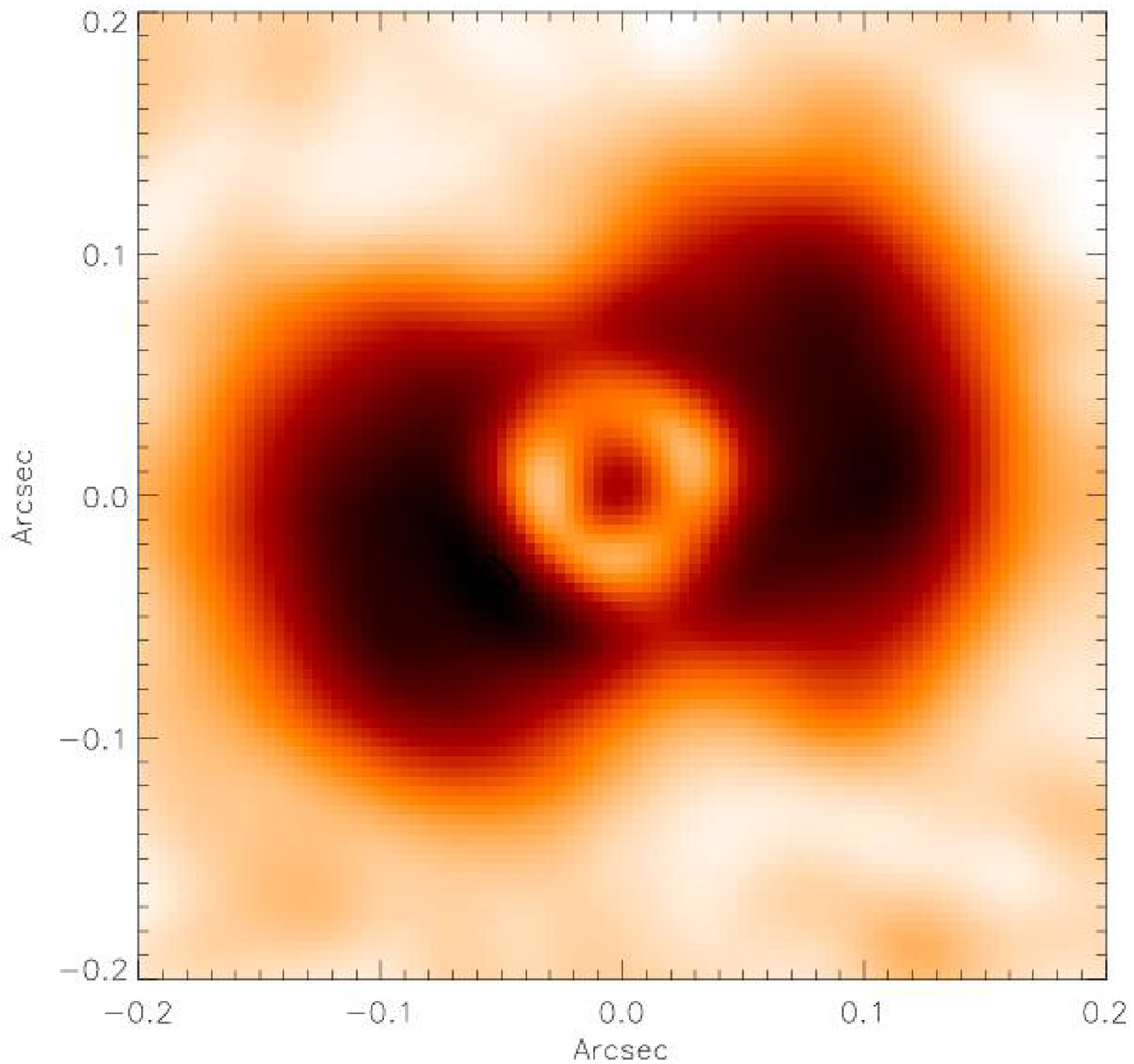}}
\caption{Bipolar fireballs everywhere! {\bf Left:} the vision of
  O.~Chesneau on RS Oph in \citep{Chesneau2012a}. {\bf Middle:} T Pyx
  bipolar fireball, as seen by AMBER and PIONIER as a face-on bipolar
  nebula \citep{Chesneau2011}. {\bf Right}: NACO/VLT 2010 K band image
  of V1280 Sco, revealing a bipolar dusty nebula
  \citep{Chesneau2012}.}
\label{nova}
\end{figure}

The enthusiasm of O.~Chesneau was only impaired by the difficulty to
interpret this extremely sparse data with a somewhat physical
model. As pragmatic as he could be, he decided to perform a simple
analysis with simple geometric models developed by N. Nardetto. He
could show this way that the fireball was not spherical (see
Fig.~\ref{nova}, left), already in the early stages of the explosion,
and that it was indeed a fireball \citep{2007A&A...464..119C}. The
lack of time series in the AMBER data was a big frustration for him as
he could only fulfill one of his two main objectives, i.e. get the
shape of the fireball but not the distance.

He later could interpret AMBER/VLTI and PIONIER/VLTI data of another
nova: T Pyx. Indeed, the broadband PIONIER/VLTI data were consistent
with a round fireball, but the spectroscopically resolved AMBER data
enabled O.~Chesneau to resolve the third dimension, providing evidence
that the nova was indeed bipolar in the very early stages of the
explosion, but seen almost exactly pole-on (see Fig.~\ref{nova},
middle). A complete interpretation was published using a sophisticated
phenomenological model that was developed by A. Meilland.

This new point of view on the system could provide an explanation to
the stunning observed slow motion HST knots around the symbiotic
system, if they were situated in an on-sky equatorial over density
\citep{Chesneau2011}. But again for T Pyx, the difficulty of
scheduling telescope time in a continuous manner was a burden to
perform the full study of the system as a function of time, and no
clear coherent time-dependent view could be used to constrain the
distance to the system.

\section[Novae]{Witnessing the start-up and evolution of a dust factory: V1280 Sco}

The study of dust around novae was made possible by the explosion in
2007 of the dust producing nova V1280 Sco. As soon as it exploded, it
appeared to be a slow nova, as it took 10 days to reach its brightness
maximum. O.~Chesneau decided to monitor this object right-away, with
AMBER/VLTI and MIDI/VLTI observations. This time, the DDT committee
was not together and the process took the regular time to go for
proposal acceptance, but O.~Chesneau did not know that the process was
1 to 2 weeks long, and therefore was extremely anxious of not getting
any data on this very interesting nova. The proposal was finally
accepted and he obtained data as soon as day 23 after the explosion,
till day 145. As the nova produced dust, its optical and near-infrared
brightness decreased quickly, making AMBER observations soon
unfeasible. It was thus important to monitor it photometrically in the
infrared to make the most out of the MIDI/VLTI observation. This
happened to be possible via a very active collaboration with Dipankar
Banerjee and his colleagues from Mt Abu in India. The VLTI follow-up
was hectic as O.~Chesneau had to permanently adapt his observing
strategy with this ever-changing object, and the observatory had
difficulties to catch up in their observing planning.

\begin{figure}
\centering
\frame{\includegraphics[width=7cm]{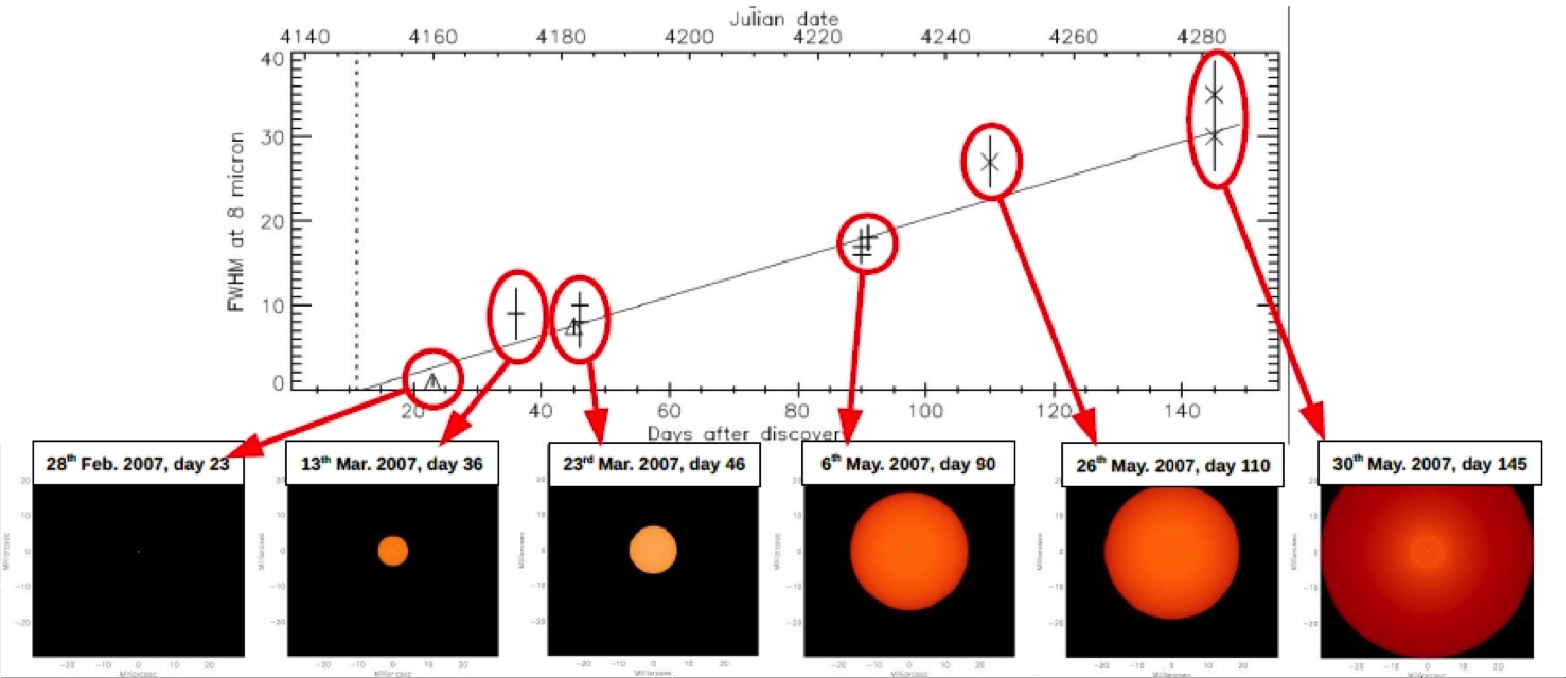}}
\frame{\includegraphics[width=5cm]{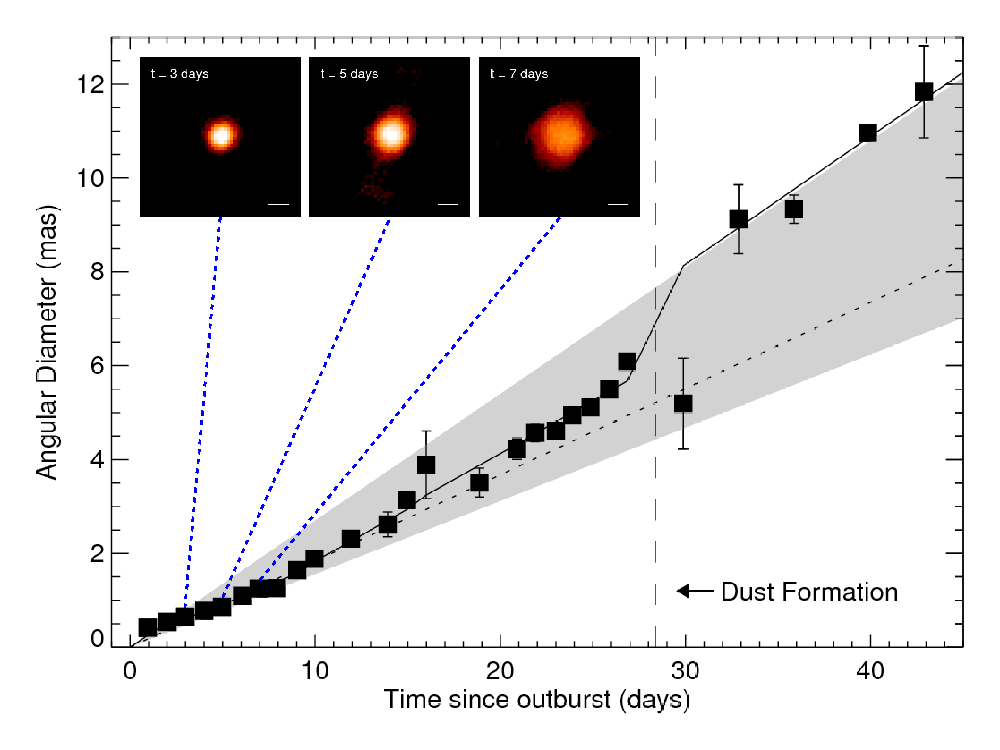}}
\caption{{\bf Left:} Evolution of the size of the circumstellar dust
  shell around V1280 Sco as seen by MIDI/VLTI, assuming the shell to
  be spherical \citep{2008A&A...487..223C}. {\bf Right:} The fireball
  expansion of Nova Delphinus 2013 witnessed by the CHARA array
  \citep{Schaefer2014b}.}
\label{novaexp}
\end{figure}

The MIDI observations immediately revealed that the dust shell was
expanding (Fig.\,\ref{novaexp}, left) and, using radiative transfer
modelling, O.~Chesneau showed that it was producing large amounts of
dust \citep{2008A&A...487..223C}. This work was made assuming that the
expanding shell was spherical, due to the sparse UV coverage of the
observations. Again, O.~Chesneau was frustrated by these observations
because he could only get a rough distance estimate, knowing that the
geometry of the fireball was wrong (as some MIDI data were showing odd
behavior compared to a round fireball), but anyway these results had
some echo following a press release from ESO.

To further study the morphology of the ejecta of V1280 Sco,
O.~Chesneau decided to use direct imaging techniques. Indeed, as the
nova shell was expanding, its angular dimension became appropriate to
be directly resolved by the VLT. He made use of the adaptive optics
imager NACO/VLT (with a resolution of $\sim$50 mas in the K band) and
the mid-infrared imager VISIR/VLT (with a spatial resolution of $\sim$
300 mas at 10$\mu$m). He secured observations in 2009, 2010 and 2011,
from t=877\,days after discovery until t=1664\,days. These
observations were strikingly spectacular, and O.~Chesneau's enthusiasm
was communicative every time he was receiving new data. An expanding
hourglass dusty shell could be monitored almost in real time (see
Fig.~\ref{nova}, right)! The nebula more than doubled in size between
July 2009 and July 2011 and we inferred a mean expansion rate of
$\sim$\,0.39 millarcsec per day in the direction of the major
axis. Most of the dust mass appears to be in the polar caps, implying
that the mass loss was predominantly polar \citep{Chesneau2012}.

\section[Novae]{Getting direct and accurate distances to Novae explosions}

All the previous attempts were missing one of the time-variable or
geometry information, or, when both were available for T Pyx, they had
a too low quality to be of much use.

The last nova O.~Chesneau could observe was Nova Delphini with several
CHARA instruments. Although he was already quite sick, he was involved
heavily in the observations preparation and follow-up. This particular
interferometric campaign was the first one making possible to get both
the geometry and time-variations information in a self-sustained data
set (see Fig.\,\ref{novaexp}, right), leading to a nice paper
\citep{Schaefer2014b} where both the geometry and an accurate distance
estimate could be obtained.

\section[Novae]{Conclusion}

The work on novae continues after the demise of O.~Chesneau, as the
monitoring framework and collaboration network he set up still
exist. His contributions to the field have changed quite drastically
the view of novae fireballs, and theoretical works are ongoing to try
to understand why the bipolar nova fireballs seem to be ubiquitous
\citep[See e.g.][]{2013IAUS..281..195M}.

\acknowledgements{The authors would like to thank L. Rolland and
  N. Nardetto for reading through this paper and making suggestions of
  improvements.}

\acknowledgements{}


\bibliographystyle{aa}
\bibliography{Chesneau}

\end{document}